\documentclass[reprint, 
 amsmath,amssymb,
prb,
floatfix
]{revtex4-1}

\usepackage{graphicx}
\usepackage{dcolumn}
\usepackage{bm}
\usepackage{hyperref}
\usepackage{upgreek}
\usepackage{xcolor}
\usepackage[version=3]{mhchem}

\begin{document}

\title{Duality of Theories for the Electrical Double Layer in Concentrated Electrolytes}

\author{Zachary A. H. Goodwin}
\email{zac.goodwin@materials.ox.ac.uk}
\affiliation{\small	John A. Paulson School of Engineering and Applied Sciences, Harvard University, Cambridge, Massachusetts 02138, United States\normalsize}
\affiliation{\small Department of Materials, University of Oxford, Parks Road, Oxford OX1 3PH, United Kingdom\normalsize}

\date{\today}

\begin{abstract}
Understanding the electrical double layer (EDL), i.e, the distribution of electrolyte at an electrified interface, in concentrated electrolytes is important for various technologies, such as supercapacitors, batteries and electrocatalysis. Atomistic approaches offer unprecedented detail, but are too computationally expensive to exhaustively investigate the EDL of concentrated electrolytes, motivating the development of continuum theories. In these concentrated electrolytes, correlations between ions and solvents are strong, through electrostatic and specific interactions, as well as significant excluded volume effects, making the development of theories challenging. Thus far, two distinct \textit{simple} theoretical approaches to understand the EDL of concentrated electrolytes, with account of these correlations beyond mean-field, \textcolor{black}{have emerged}. One is a \textcolor{black}{more conceptual} local-density approximation (LDA), based on treating electrostatic and specific interactions beyond mean-field through ionic aggregation and solvation; where reasonable agreement with experiments in terms of integrated quantities, but poor agreement for ion profiles and Debye capacitance \textcolor{black}{is obtained}. The other approach is to treat electrostatic correlations and excluded volume effects more rigorously with beyond LDA \textcolor{black}{methods}, but at the cost of simplifying the chemical interactions between species; where excellent agreement for ion profiles, differential capacitance, etc. \textcolor{black}{can be obtained}, but mainly for the simplified hard-sphere systems that the theories are based on. Here, we describe the merits and downfalls of these two approaches, how they have contributed to understanding anomalous underscreening, and outline future directions for these theoretical approaches.
\end{abstract}

\maketitle

\section{Introduction}

Concentrated electrolytes, such as ionic liquids (ILs)~\cite{Kornyshev2007,Fedorov2014}, water-in-salt electrolytes (WiSEs)~\cite{Suo2015} and salt-in-ionic liquids (SiILs)~\cite{molinari2019general,molinari2019transport}, are increasingly being studied and finding applications in various technologies, such as batteries~\cite{Goodenough2013ThePerspective,xu2014electrolytes,li2020new}, supercapacitors~\cite{Fedorov2014} and electrocatalysis~\cite{Gebbie2023}. While the high salt concentrations of these electrolytes have some downsides, such as high viscosities and low-to-average conductivities, the stability and safety of these electrolytes often balance these issues, making them promising alternatives~\cite{Fedorov2014,Goodenough2013ThePerspective,xu2014electrolytes,Gebbie2023,Zheng2017Uni}. 

A central theme of such technologies is the interface of these electrolytes with an electrode~\cite{Fedorov2014,Gebbie2023,Zheng2017Uni,Cheng2022Sol}. On the electrolyte side, how the ions and solvent (if present) arrange at this interface is referred to the electrical double layer (EDL)~\cite{Kornyshev2007,Fedorov2014,Bazant2009a}. Understanding the EDL is of great importance in these technologies~\cite{Gebbie2023}, as it is directly related to the energy stored in a supercapacitor~\cite{Fedorov2014}, for example, and remains an active field, but still has its challenges. 

\textcolor{black}{At} high salt concentration\textcolor{black}{s}, there are strong electrostatic correlations and excluded volume effects; moreover, the ions and solvents are often complicated molecular species with significant non-electrostatic interactions~\cite{Borodin2017Mod,Borodin2014SEI}. Therefore, understanding the EDL of concentrated electrolytes has been challenging. Simulating the EDL with classical force fields in molecular dynamics (MD) is routinely performed~\cite{Zheng2017Uni,Cheng2022Sol,Qisheng23JACS}. However, the accuracy of the empirical models are often under question. On the other hand, \textit{ab initio} MD (AIMD) offers an alternative accurate route, but these methods are prohibitively expensive. Recently, machine learning interatomic potentials (MLIPs) have offered to bridge the cost-accuracy gap between these methods~\cite{batzner2021se,NEURIPS2022_4a36c3c5,Musaelian2023,Deringer2019}. However, long-range corrected MLIPs are still being developed, and local MLIPs remain orders of magnitude more expensive than empirical potentials. Therefore, there is still motivation to develop theories for the EDL of concentrated electrolytes.

Developing theories for the EDL has a long history, at least for dilute electrolytes~\cite{Bazant2009a}. For concentrated electrolytes, the first paradigm change came with introducing excluded volume in the Kornyshev model~\cite{Kornyshev2007,kilic2007a}. The theory predicted a saturation and decay in differential capacitance, but it occurred over very small voltage ranges~\cite{Goodwin2017a}. Moreover, it predict\textcolor{black}{s} monotonically decaying ion profiles, but damped oscillations at low voltages \textcolor{black}{were} known to occur, referred to as overscreening~\cite{Fedorov2014}. The Bazant-Storey-Kornyshev theory was the first to simply describe overscreening~\cite{Bazant2011,pedro2018}. Since these advancements, there has been significant effort to further include correlations beyond mean-field in an \textit{accurate and simple} way.

\begin{figure*}
    \centering
    \includegraphics[width=0.8\linewidth]{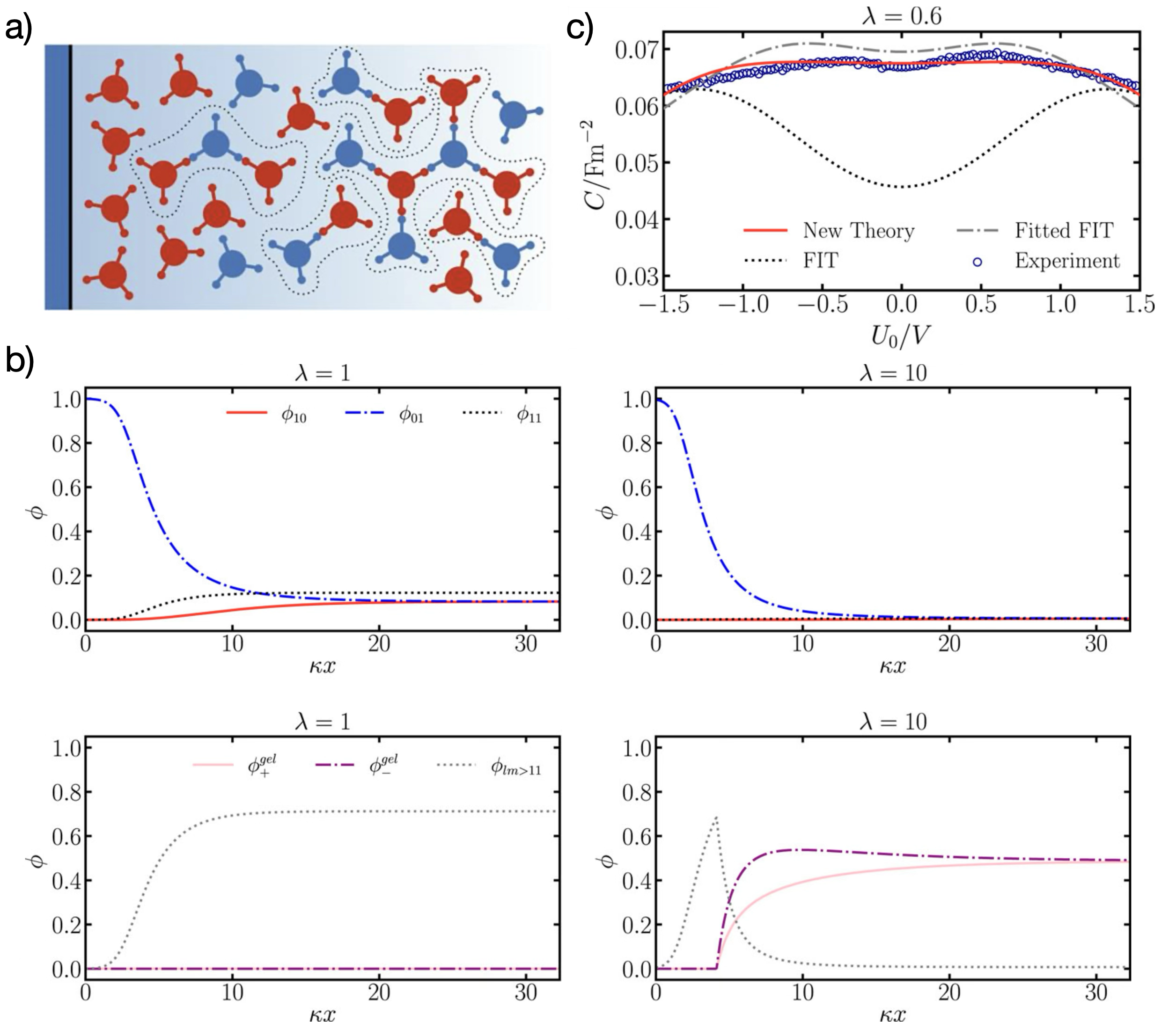}
    \caption{Summary of general trends from ionic aggregation theory in ILs. a) Schematic illustration of the EDL of ILs with consistent account of thermoreversible associations. Anions are shown in blue, cations in red. The dangling bonds represent the number of associations they can form, and aggregates are denoted by dotted lines surrounding them. Close to the interface, ionic associations are destroyed, and free species dominate. b) Volume fractions ($\phi$) of various species ($\phi_{10}$ is free cations, $\phi_{01}$ is free anions, $\phi_{11}$ is ion pairs, $\phi_{+/-}^gel$ are cations/anions in the gel, and $\phi_{lm > 11}$ are the remaining aggregates) as a function from the positively charged interface, where $\kappa$ is the inverse Debye length. In the titles are the association constants, $\lambda = \exp(-\beta \Delta f)$, where $\beta$ is inverse thermal energy and $\Delta f$ is the free energy of an association. c) Experimental differential capacitance (from Ref.~\citenum{Monchai2018}) as a function of applied voltage, compared against the ``new theory" from Ref.~\citenum{Goodwin2022EDL} with the indicated association constant, compared against free ion theories (FIT) from Ref.~\citenum{Goodwin2017a}. Figures reproduced from Ref.~\citenum{Goodwin2022EDL}, under the CC-BY-4.0 license.}
    \label{fig:1}
\end{figure*}

One recent challenge that has tested our faith in these theories is the observation of ``anomalous underscreening''~\cite{Gebbie2013,Gebbie2015}. Initially observed in surface force balance measurements with ILs, where force decay lengths of $\sim$7nm were reported, $\times100$ what would be expected from the Debye screening length~\cite{Gebbie2013,Gebbie2015,Han2020}. Later, this was generalized over a range of concentrations and electrolytes~\cite{smith2016electrostatic,perez2017underscreening,Han2021WiSE,Zhang2024}. These observations have stimulated the development of EDL theories in concentrated electrolytes~\cite{hartel2023anomalous,Kjellander2016,avni2020charge,adar2019screening,Pedro2020,Goodwin2022EDL}.

It was suggested these long decay lengths could arise from ion pair formation, or from these ion pairs aggregating into a network~\cite{Gebbie2013,Gebbie2015,Han2020}. The inclusion of ionic aggregates is a way of treating correlations beyond mean-field electrostatics~\cite{Goodwin2017a,Zhou2022Agg}, but a consistent theory to include them in the EDL has only recently been formulated~\cite{Goodwin2022EDL,Goodwin2022Kornyshev}. While treating correlations through aggregate formation is one approach, another approach is to explicitly treat electrostatic correlations~\cite{Girotto2018,Frydel2012,Girotto2017,Pedro2020,Bui2025}. Therefore, in recent years, there has been a duality of approaches taken to understand the EDL of concentrated electrolytes. In what follows, we outline the progresses for these different approaches. After discussing their relative successes and limitations, we will discuss possible future directions, before highlighting additional works which include effects beyond these simple electrolyte theories.

\section{Ionic Aggregation Approach}

\begin{figure*}
    \centering
    \includegraphics[width=0.8\linewidth]{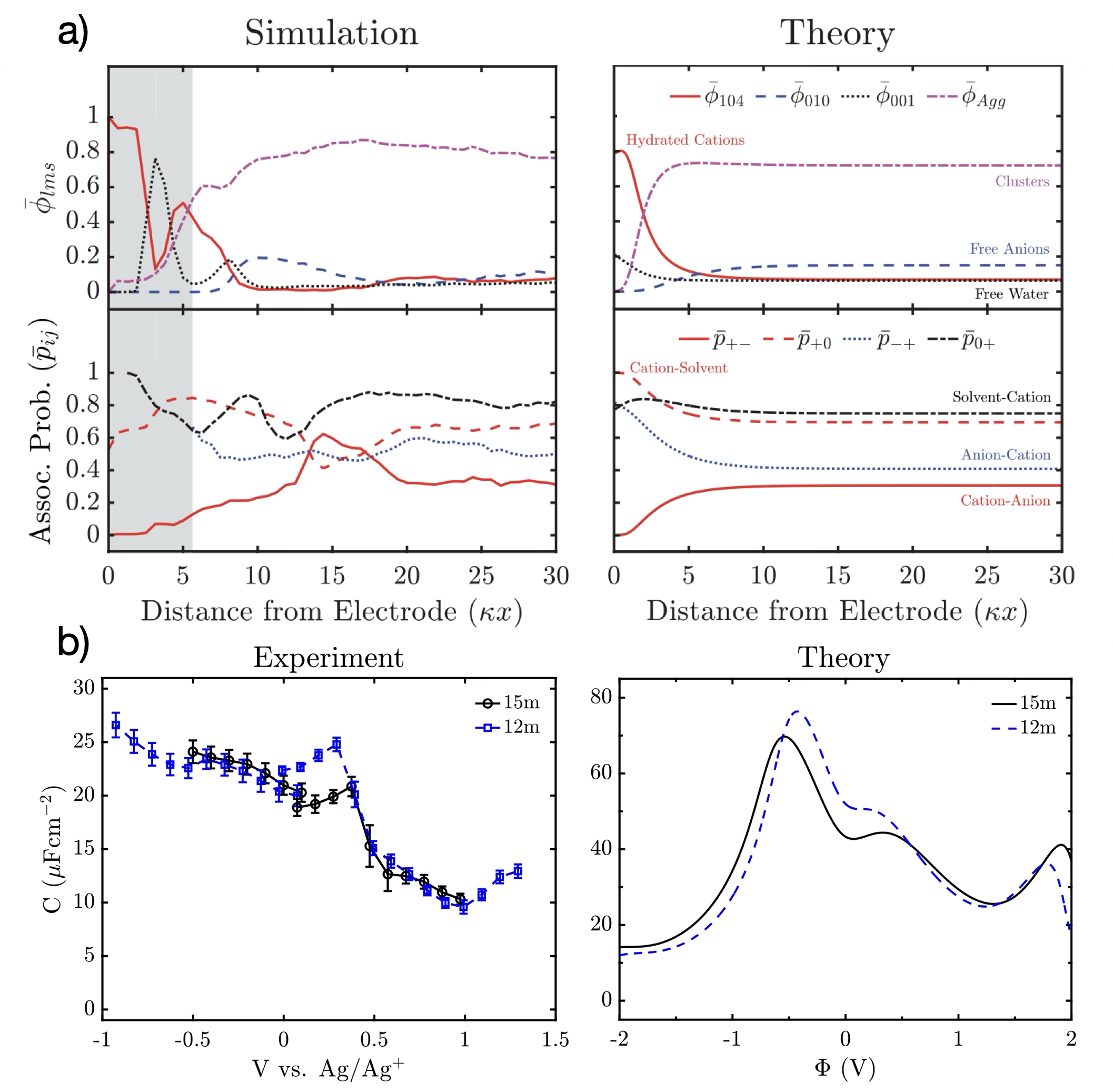}
    \caption{Overview of ionic aggregation approach in describing the EDL of LiTFSI WiSE. a) Comparison between simulation and theory for 15~m LiTFSI WiSE at a surface charge of -0.2 Cm$^{-2}$. Top panels show how the volume fractions \textcolor{black}{($\phi$)} of various aggregates, as indicated, vary within the EDL. The bottom panels show how association probabilities between the species change within the EDL. The gray region denotes where interfacial interactions become important, b) Comparison between experimental and theoretical differential capacitance of 15m and 12m LiTFSI WiSE. \textcolor{black}{The experimental differential capacitance curve is referenced to Ag/Ag$^+$, while the theory is referenced to the potential of zero charge (PZC). In the experiments, the PZC is often taken to be the first minimum~\cite{zhang2020potential}, which is just above 0 V at 15~m and just below 0 V at 12~m.} Figures reproduced from Ref.~\citenum{Markiewitz2025}, under the CC-BY-4.0 license.}
    \label{fig:2}
\end{figure*}

As previously mentioned, a concept to include correlations beyond mean-field is through introducing ion pairs and ionic aggregates as distinct species~\cite{Gebbie2013,Gebbie2015,Han2020}. In WiSEs and SiILs these concepts are ubiquitous~\cite{mceldrew2021ion,McEldrewsalt2021}, although in ILs it is less well accepted~\cite{mceldrew2020corr,Welton2025}. Until 2020, ion pairs were only included in these descriptions, or not even explicitly, with free ion approaches ignoring any information of the aggregates~\cite{Goodwin2017a}. In concentrated electrolytes, however, this is not conceptually appealing, since in such a dense fluid higher order interactions between ion pairs and aggregation of them would be expected~\cite{mceldrew2020theory}. This was noted by Gebbie \textit{et al.} in their interpretation of the long force decay lengths~\cite{Gebbie2013,Gebbie2015,Han2020}. McEldrew \textit{et al.} were the first to introduce a thermodynamic theory that described the formation of all possible ionic aggregates, provided they were Cayley-trees, with thermoreversible associations, and the existence of a percolating ionic network~\cite{mceldrew2020theory,mceldrew2020corr,mceldrew2021ion,McEldrewsalt2021}.

Within a local-density approximation, i.e., where the variables of the theory only depend on position (not derivatives of variables, or integral quantities), Ref.~\citenum{Goodwin2022EDL} extended this formalism to predict the EDL of ILs, schematically shown in Fig.~\ref{fig:1}a), with consistent account of \textit{all} chemical equilibria. The ion profiles from this theory can be seen in Fig.~\ref{fig:1}b). Consistent with previous approaches, electric field destruction of ionic aggregates and ionic networks \textcolor{black}{was found}~\cite{Goodwin2017a}. Surprisingly, the gel phase (percolating ionic network) \textcolor{black}{was found to} screen the electrode since it becomes charged~\cite{Goodwin2022EDL}. From the destruction of the gel, ionic aggregates can \textcolor{black}{intermittently} increase. The predicted capacitance also agreed more closely with experiments than previous free ion approaches, as seen in Fig.~\ref{fig:1}c). This approach, however, was not able to reconcile ionic networks with ``anomalous underscreening''~\cite{Gebbie2013,Gebbie2015,Han2020}.

\begin{figure*}
    \centering
    \includegraphics[width=0.8\linewidth]{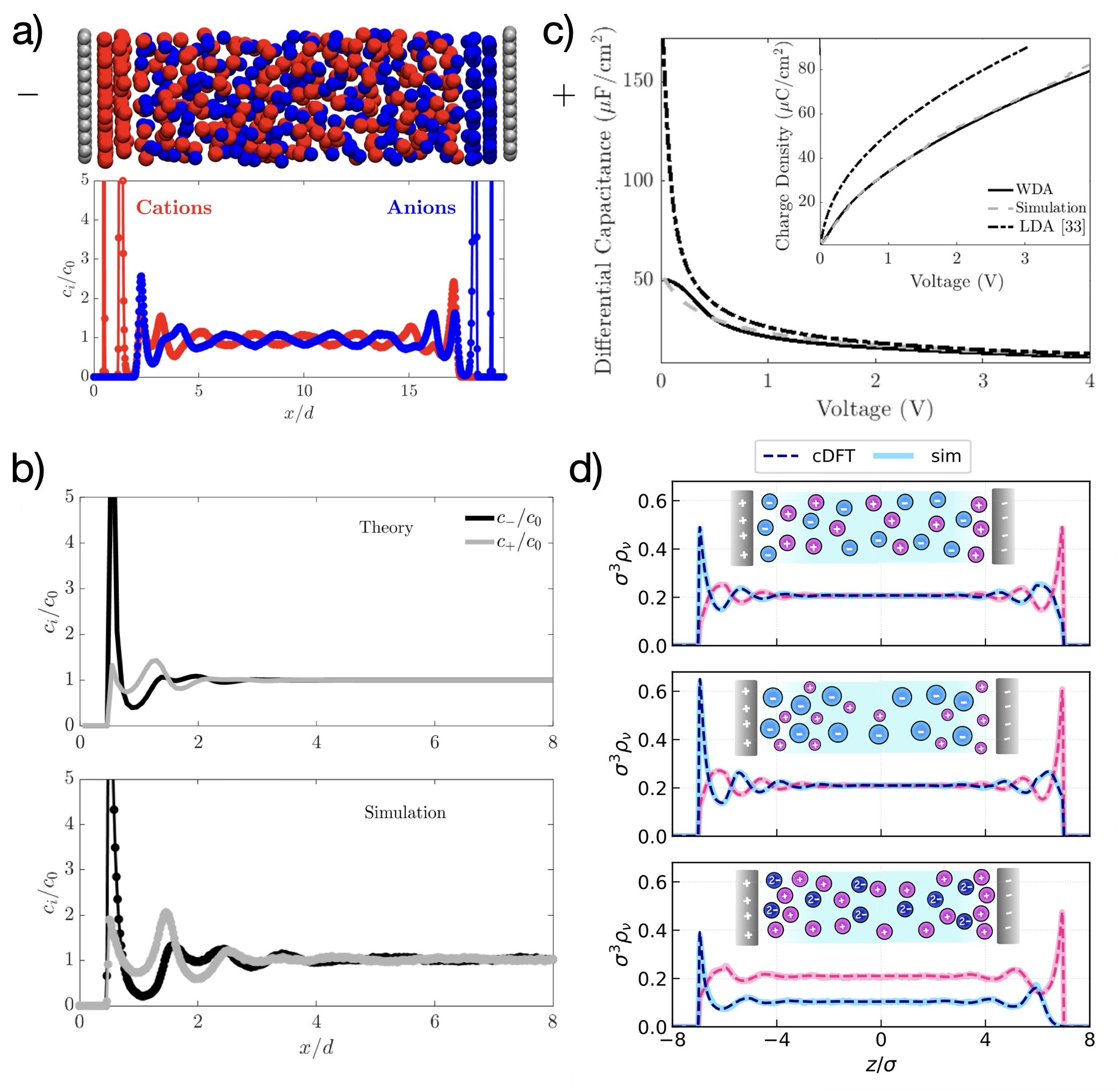}
    \caption{Summary of recent direct ion-interaction approaches, and success at predicting ionic layering. a) MD simulation of charged Lennard-Jones spheres, with cations in red and anions in blue (top panel). Concentration of cations and anions ($c_i$, red and blue, respectively) over bulk concentration $c_0$, as a function of distance from the electrodes, relative to the ion diameter $d$ (bottom panel). Surface charge of 0.12 Cm$^{-2}$ at a temperature of 300~K. b) Ion layering profiles at a 0.01 Cm$-2$ charged surface from theory and simulations, where remarkable agreement between them is found. c) Differential capacitance as a function of applied voltage for the WDA of Ref.~\citenum{Pedro2020}, compared against the simulations, which match well, and the LDA theory of Refs.~\citenum{Kornyshev2007,kilic2007a}, which predicted much larger values. a)-c) have been reproduced from Ref.~\citenum{Pedro2020} with permission. d) Ion profiles from the classical-DFT with neural network learned correlation function, compared to classical simulations of the electrolytes. Examples of charge and size symmetric, charge symmetric but size asymmetric, and size symmetric but charge asymmetric electrolytes simulations all show remarkable agreement with the developed cDFT. Figure reproduced from Ref.~\citenum{Bui2025} with the CC-BY-4.0 license.}
    \label{fig:3}
\end{figure*}

This formalism was further developed by Markiewitz \textit{et al.} to work for asymmetric, multicomponent electrolytes, such as SiILs~\cite{markiewitz2024,Zhang2024} and WiSEs~\cite{Markiewitz2025}. For WiSEs, a thorough comparison against MD simulations and experimental measurements was achieved, where qualitative agreement was found for the diffuse EDL as seen in Fig.~\ref{fig:2}a), with the largest discrepancy occurring right at the interface. To understand why there was a break down of the theory right at the interface, in the Helmholtz layer, Ref.~\citenum{GoodwinHelm2025} analysed in detail the solvation distributions of a battery electrolyte, and found that the maximum number of solvents an ion can associate to reduces at the interface, since the electrode blocks and interacts with the ion \textcolor{black}{association sites}. \textcolor{black}{Moreover, the predicted differential capacitance, see Fig.~\ref{fig:2}b), agrees qualitatively with experimental measurements, but the theory over-predicts the magnitude of the capacitance owing to the employed charge rescaling parameter, lack of Stern layer and the LDA nature of this theory~\cite{Markiewitz2025}.}

\section{Ionic Interaction Approach}

On the other hand, approaches which more rigorously treat non-local electrostatics and excluded volume effects have existed for quite some time, which overcome the deficiencies of simple LDA approaches~\cite{Girotto2018,Frydel2012,Girotto2017,Pedro2020,Bui2025}. Charged hard sphere models (such as the restricted primitive model) have also been extensively studied for concentrated electrolytes~\cite{hartel2023anomalous}. The solution of such approaches can be difficult, however, and applying these methods to electrokinetics or other phenomena is challenging. Recently, however, approximations of such approaches have been able to attain a similar accuracy, but at an ease of implementation and solution~\cite{Pedro2020,Bui2025}.

One such example is the so-called weighted density approximation (WDA) of de Souza \textit{et al.}~\citenum{Pedro2020}. At linear response, a fourth-order modified Poisson-\textcolor{black}{Boltzmann} equation was derived, with a screening length of $\lambda_D/\lambda_s = (1 \pm \sqrt{1 - 4(d/\sqrt{24}\lambda_D)^2})/2(d/\sqrt{24}\lambda_D)^2$, where $\lambda_D$ is the Debye length and $d$ is the ion diameter. Clearly, this was also not able to rationalize the ``anomalous underscreening'' paradox, but unlike the BSK~\cite{Bazant2011} approach the screening length is \textcolor{black}{directly} tied to the ion diameter, $d$~\cite{Pedro2020}. As seen in Fig.~\ref{fig:3}a)-c), its predicted ionic layering and differential capacitance agreed remarkably well against simulations of charged Lennard-Jones spheres. A key advantage of this approach is that it is not significantly \textcolor{black}{more} difficult to solve than LDA modified Poisson-Boltzmann approaches, making it attractive to apply to complicated processes, and systems with water~\cite{pedro2022polar}.

The WDA approach essentially \textcolor{black}{has} no free parameters~\cite{Pedro2020}\textcolor{black}{, meaning its predictions are robust provided the assumptions of the theory are valid}. With the system of equations set up to describe charged hard spheres, applying it to more realistic electrolytes causes disagreements. To overcome such issues, and have a more general formalism for learning the excess chemical potential, Bui and Cox~\cite{Bui2025} have developed neural networks to learn how the one-body direct correlation functions depends on density, and including electrostatics in a mean-field way. To generate data to train the neural network, $\sim$2500 grand canonical simulations were required~\cite{Bui2025}. While this is seemingly a lot, the advantage of this approach is that different force fields can be used to then have a theory which \textcolor{black}{can} be systematically tuned, without further approximations. As seen in Fig.~\ref{fig:3}d), the resulting ion profiles at charged interfaces agree remarkably well with the simulations. Moreover, this approach can also be applied to charge asymmetric and size asymmetric systems, and is relatively easy to implement with more complicated processes, since this method is not significantly more complicated than an LDA approach. 

\section{Conclusion and Outlook}

In recent years, two distinct approaches have emerged in the domain of simple theories for concentrated electrolytes at electrified interfaces. On one side, \textcolor{black}{there is} the approach of treating electrostatic correlations beyond mean-field (and specific interactions) through the formation of \textcolor{black}{thermoreversible} associations~\cite{Goodwin2022EDL}. This approach has had some success in explaining differential capacitance curves and aggregation length scales of real electrolytes, such as LiTFSI in water~\cite{Markiewitz2025}. However, \textcolor{black}{while analytically simple and conceptually tractable}, since it is based on an LDA functional, the ionic profiles and absolute values of differential capacitance do not quantitatively agree with experiments and simulations without additional fitting parameters. Moreover, it also has not been able to rationalize the ``anomalous underscreening'' observations, although it has hinted to new directions of their explanations, i.e., confinement induced aggregation and elastic responses from the ionic networks~\cite{Zhang2024}. On the other hand, there have recently been relatively simple approaches which treated electrostatics and excluded volume effects more rigorously~\cite{Pedro2020,Bui2025}. These approaches have excellent agreement with ionic profiles from simulations and differential capacitance\textcolor{black}{, but as the approaches have no free fitting parameters and assumed charged spheres, they struggle to generalise to more realistic electrolytes}. \textcolor{black}{Finally,} the machine learning approach of Bui and Cox is flexible \textcolor{black}{enough} to include more details of the underlying force field in the theory~\cite{Bui2025}, it remains to be tested in electrolytes with strong specific interactions, such as WiSEs.

In the future, the challenge will be to unify these distinct approaches, such that a theory that can have chemically specific interactions and aggregation/solvation, but also rigorous electrostatics and excluded volume effects. This is not only a technical challenge, but a conceptual one. As overscreening is a representation of ionic aggregation~\cite{Goodwin2022EDL,avni2020charge}, one must be careful with how to include the correlations to not double count interactions, and it is not clear which side will provide the best starting point for this (aggregation vs correlation).

For understanding ``anomalous underscreening", the theory of de Souza \textit{et al.}~\cite{Pedro2020} predicted an exponent of 2, which is in agreement with a large body of other theories~\cite{adar2019screening,Espinosa2023rev}. Therefore, if the experiments are in equilibrium and there is no influence from the surfaces, these theories demonstrate that the explanation cannot originate solely from electrostatic correlations. Recently it was demonstrated ``anomalous underscreening" could be largely from hydrodynamic interactions~\cite{Cross2026}, not a bulk electrostatic effect \textcolor{black}{at equilibrium}. \textcolor{black}{This can also rationalize why long-ranged forces were not observed in AFM with atomically sharp tips, and until recently~\cite{Tilger2026} why they were not observed colloidal probe AFM measurements~\cite{kumar2022absence,Espinosa2023rev}}.

Beyond these theoretical approaches, there are emerging trends in the theoretical community that will contribute to the development of these methods. The approaches discussed so far have ignored practically all interactions with the interface~\cite{GoodwinHelm2025}.  Electrochemistry, however, requires us to understand how ions and electrons behave at the interface between electrodes and electrolytes. Therefore, to further understand this, theoretical approaches which treat ions and electrons on the same footing will provide insight into correlations between them across the interface~\cite{Hedley2025,Peng2023eld,Bruch2024,Wang2025DPFT}. 

\section{Acknowledgments}

\textcolor{black}{I thank J P de Souza, Q Zheng and R Espinosa-Marzal for helpful discussions}. Z.A.H.G acknowledges support through the Glasstone Research Fellowship in Materials and The Queen's College, University of Oxford.

\bibliography{WiSE}

\end{document}